\documentclass{article}

\usepackage{amsmath,amssymb,amsthm}
 \DeclareMathOperator{\tr}{Tr}
\DeclareMathOperator{\1}{\textbf{1}}
\DeclareMathOperator{\spa}{span}
\DeclareMathOperator{\I}{\textbf{1}}
\DeclareMathOperator{\II}{\textbf{1}_4}

\newcommand{\ot}{\otimes}
\newcommand{\Ix}{\sigma_x}
\newcommand{\Iy}{\sigma_y}
\newcommand{\Iz}{\sigma_z}

\newtheorem{theorem}{{\large T}HEOREM}[section]

\newtheorem{proposition}{{\large P}ROPOSITION}[section]
\newtheorem{example}{{\large E}XAMPLE}[section]
\newtheorem{definition}{{\large D}EFINITION}[section]

\theoremstyle{definition}
\newtheorem{remark}{{\large R}EMARK}[section]

\date{}

\title{A General  Framework for Recursive Decompositions of Unitary Quantum Evolutions}

\author{Mehmet Dagli\thanks{\ \
Electronic address: mdagli@iastate.edu}\,,\ \ Domenico
D'Alessandro\thanks{\ \
Electronic address: daless@iastate.edu}\,, and Jonathan D.H. Smith \thanks {\ \
Electronic address: jdhsmith@iastate.edu}\\[2ex]
Department of Mathematics, Iowa State University, Ames IA 50011,
USA}

\begin{document}

\maketitle


\begin{abstract}

Decompositions of the unitary group $U(n)$ are useful tools in
quantum information theory as they allow one to decompose unitary
evolutions into local evolutions and evolutions causing
entanglement. Several recursive decompositions have been proposed in
the literature to express unitary operators as products of simple
operators with properties relevant in entanglement dynamics. In this
paper, using the concept of grading of a Lie algebra, we cast these
decompositions in a unifying scheme and show how new recursive
decompositions can be obtained. In particular, we propose  a new
recursive decomposition of the unitary operator on $N$ qubits, and we
give a numerical example.

\end{abstract}


\vspace{0.25cm}



\section{Introduction}

Decompositions of the unitary Lie group $U(n)$ serve to factorize
any element $X_f \in U(n)$ as a product $X_f=X_1X_2 \cdots X_m$,
where $X_1,\dots,X_m$ are (elementary) factors in $U(n)$.  There are
several reasons to study such decompositions for unitary evolutions
in quantum mechanics. They allow one to analyze the dynamics of a
quantum system in terms of simpler, possibly meaningful, factors. In
particular, for multipartite systems they allow the identification
of the local and entangling parts of a given evolution. In this
context one can study entanglement dynamics \cite{SG}, \cite{SGD},
\cite{Zhang}. From a more practical point of view, they allow one to
decompose the task of designing a given evolution, such as a quantum
gate, into simpler, readily available dynamics (cf., e.g.,
\cite{Lloyd}). In particular, in multipartite systems few entangling
evolutions are typically available. Lie group decompositions  are
also useful in control problems \cite{BOOK} and in the solution of
some algebraic problems of interest in quantum information
\cite{Petz}. For these reasons several decompositions have been
introduced in recent years \cite{SG}, \cite{SGD}, \cite{DF},
\cite{DR}, \cite{NS}. In \cite{SG}, \cite{SGD}, a decomposition
called the {\it Concurrence Canonical Decomposition (CCD)} was
studied in the context of entanglement theory. The CCD is a way to
decompose every unitary evolution on $N$ q-bits into a part that
does not modify the concurrence on the $N$ q-bits, and a part that
does. It is a Cartan decomposition in that it corresponds to a
symmetric space of $SU(2^{N})$ \cite{Helgason}.  In \cite{DF} the
CCD was further studied and generalized to multipartite systems of
arbitrary dimensions. The resulting decomposition was called an {\it
Odd-Even Decomposition (OED)}. The OED is a decomposition of unitary
evolutions on multipartite systems constructed in terms of
decompositions on the single subsystems. {\em Recursive}
decompositions such as the ones in \cite{DR} and \cite{NS}
recursively apply the Cartan decomposition theorem in order to
decompose the factors into simpler ones.

The present paper is devoted to recursive decompositions. Using the
relation between Cartan decompositions of Lie algebras and Lie
algebra gradings, we show that the recursive decompositions of
\cite{DR} and \cite{NS} are a special case of a general scheme from
which several other recursive decompositions can be obtained.

The paper is organized as follows. Most of the content of section
\ref{BM} is background material concerning the basic concepts of
Cartan decompositions of Lie groups and algebras, with particular
emphasis on decompositions of $U(n)$. We also describe the main
ingredients of the CCD decomposition of \cite{SG} and \cite{SGD};
the OED decomposition of \cite{DF}; and the recursive decompositions
of Khaneja and Glaser \cite{NS}, and D'Alessandro and Romano
\cite{DR}. One extension of the procedure used for the OED
decomposition is presented in Theorem \ref{AIIIOED}. In section
\ref{LAG}, we describe gradings of Lie algebras, and establish a
link between gradings and recursive decompositions. This gives a
general method to develop recursive decompositions of $U(n)$. We
show in section \ref{Special} how the recursive decompositions of
\cite{DR} and \cite{NS} are special cases of this general procedure,
and how new recursive decompositions can be obtained. In section
\ref{CIE}, we give a numerical example illustrating the calculation
of the recursive decompositions described in section \ref{Special}.
Some concluding remarks are presented in section \ref{conc}.


\section{Cartan decompositions of the unitary group}
\label{BM}

\subsection{Cartan decompositions of a Lie Algebra}
\label{BMLAG}

A \emph{Cartan decomposition} of a semisimple Lie algebra
$\mathcal{L}$ is a vector space decomposition
\begin{equation}
\label{CartanDecomposition}
\mathcal L = \mathcal K \oplus \mathcal P \,,
\end{equation}
where the subspaces $\mathcal{K}$ and $\mathcal{P}$ satisfy the
commutation relations
\begin{equation*}
[\mathcal{K},\mathcal{K}] \subseteq \mathcal{K}, \quad
[\mathcal{K},\mathcal{P}] \subseteq \mathcal{P}, \quad
[\mathcal{P},\mathcal{P}] \subseteq \mathcal{K}\,.
\end{equation*}
The pair $(\mathcal K,\mathcal P)$ is called a \emph{Cartan pair} of
$\mathcal L$. In particular, $\mathcal{K}$ is closed under the Lie
bracket, and is therefore a Lie subalgebra of $\mathcal{L}$. A
Cartan decomposition of a Lie algebra $\mathcal{L}$ induces a
decomposition of the connected Lie group associated to
$\mathcal{L}$, which we denote by $e^{\mathcal{L}}$. In particular,
every element $L$ of $e^{\mathcal{L}}$ can be written as
\begin{equation}
\label{KPDeomposition}
L=KP\,,
\end{equation}
where $K\in e^{\mathcal{K}}$ and $P$ is the exponential of an
element in $\mathcal{P}$. Since
$[\mathcal{P},\mathcal{P}]\subseteq\mathcal{K}$, any Lie subalgebra
contained in $\mathcal{P}$ is necessarily Abelian. A maximal Abelian
subalgebra $\mathcal{H}$  contained in $\mathcal{P}$ is called a
\emph{Cartan subalgebra}, and the common dimension of all the maximal Abelian subalgebras $\mathcal{H}$ is
called  the \emph{rank} of the decomposition. Indeed, although the Cartan subalgebra
is not unique, it may be shown that  two Cartan
subalgebras $\mathcal{H}$ and $\mathcal{H}_1$ are conjugate via an
element of $e^{\mathcal K}$. This means that  there exists $S\in
e^{\mathcal{K}}$ such that $\mathcal{H}=Ad_{S}(\mathcal{H}_1)$. Here
$Ad_{S}$ denotes the adjoint map defined as $Ad_S(H):=SHS^\dagger$
for  $H \in {\mathcal{L}}$.

Let $\mathcal{H}$ be a Cartan subalgebra of $\mathcal{L}$. One can
prove that
$$
{\mathcal{P}}={\tiny\bigcup_{S\in e^{\mathcal{K}}}}Ad_{S}({\mathcal{H}}) \,,
$$
and therefore
$$
\exp( \mathcal P ) = \bigcup_{ S \in e^{ \mathcal K }} Ad_{S}( e^{\mathcal H } ).
$$
It follows that $P$ in (\ref{KPDeomposition})
has the form $ P=SAS^\dagger $, with $ S\in e^{\mathcal K} $ and  $
A\in e^{\mathcal{H}} $. Hence, from (\ref{KPDeomposition}),  each
$L\in e^{\mathcal L}$ can be written as
\begin{equation} \label{KAK2}
L=K_1AK_2\,,
\end{equation} where $K_1,K_2 \in e^{\mathcal{K}}$ and
$A\in e^{\mathcal{H}}$. This decomposition is known as  the
$KAK$ decomposition of the Lie group $e^{\mathcal L}$.

Cartan classified all the Cartan decompositions of the classical
Lie algebras \cite{Helgason}. In particular, up to conjugacy, there
exist three types of Cartan decomposition of the special unitary
Lie algebra $\mathfrak{su}(n)$, the Lie algebra of
skew-Hermitian matrices with zero trace. The decompositions  are
classified as AI, AII, and AIII.

A decomposition of type AI is the Cartan decomposition of
$\mathfrak{su}(n)$ into purely real and purely imaginary matrices,
i.e.,
\begin{equation}
\label{CD-AI} \mathfrak{su}(n) = \mathfrak{so}(n) \oplus
\mathfrak{so}(n)^\perp.
\end{equation}
The orthogonality is given by the inner product $\langle A\,,B
\rangle=\tr(AB^\dagger)$ where $A,B\in\mathfrak{su}(n)$. The
diagonal matrices in $\mathfrak{so}(n)^\perp$ span a maximal Abelian
subalgebra, so the rank of the decomposition is $n-1$.

A decomposition of type AII is of the form
\begin{equation}
\label{CD-AII}
\mathfrak{su}(2n) = \mathfrak{sp}(n) \oplus
\mathfrak{sp}(n)^\bot\,,
\end{equation}
where   $\mathfrak{sp}(n)$ is the Lie algebra of symplectic matrices,
namely the  subalgebra of $\mathfrak{su}(2n)$ of matrices $A$
satisfying
$$
AJ+JA^T=0,
$$
in which $J$ is the $2n \times 2n$ matrix
$$
J:=\begin{pmatrix}0 & {\bf 1}_n \\
 -{\bf 1}_n &
0\end{pmatrix}.
$$
Here and in the rest of this paper, we denote by ${\bf 1_n}$ the $n\times n$ identity matrix. The rank
of the decomposition AII is again $n-1$.

A decomposition of type AIII is defined in terms of two positive
integers $p$ and $q$ with $p+q=n$. The decomposition is
\begin{equation}
\label{AIIIdec}
 \mathfrak {su}(n):={\cal K} \oplus {\cal P},
\end{equation}
where ${\cal K}$ is spanned by block diagonal matrices
\begin{equation}
\label{bloF}
F:=\begin{pmatrix}X_{p\times p} & 0 \\
0 & Y_{q \times q}\end{pmatrix},
\end{equation}
  with $X_{p \times
p}$ and $Y_{q \times q}$ skew-Hermitian and $\tr(X_{p \times
p})+\tr(Y_{q \times q})=0$. The rank of this decomposition is $\min
\{p,q\}$.

Each Cartan decomposition of $\mathfrak{su}(n)$ is conjugate to
one of the decompositions of type AI, AII, and AIII. In other words, if
$\mathfrak{su}(n)={\cal K} \oplus {\cal P}$ is a Cartan
decomposition of $\mathfrak{su}(n)$,  there exists a unitary matrix
$T$ such that $\mathfrak{su}(n)=T{\cal K}T^\dagger \oplus T{\cal
P}T^\dagger$ is in one of the forms AI, AII and AIII. These
decompositions can be expressed in forms of interest in various
contexts, for example with matrices expressed as tensor products of
operators on single subsystems in a multipartite quantum system.

In the following, we shall find it convenient to extend these
decompositions to decompositions of $\mathfrak{u}(n)=
\mathfrak{su}(n) \oplus \spa \{i \1_n \}$, the Lie algebra of
$U(n)$. Consider a Cartan decomposition of the special unitary Lie
algebra $\mathfrak{su}(n)$ of type either AI or AII. Since the
identity matrix $\1_n$ commutes with each element of
$\mathfrak{su}(n)$, the Cartan decompositions of $\mathfrak{su}(n)$
of types AI (\ref{CD-AI}) and AII (\ref{CD-AII}) can be naturally
extended to decompositions of $\mathfrak{u}(n)$ by replacing
$\mathcal P$ with $\mathcal P\oplus \spa \{ i\1_n \}$. We also
denote these decompositions of types AI and AII. In both Cartan
decompositions, the rank becomes $n$. For decompositions of type
AIII, we find it convenient to include $\text{span} \{i \1_n \}$ in
the Lie algebra part, and replace ${\cal K}$ with ${\cal K} \oplus
\text{span} \{i \1_n \}$, so as to lift the restriction $\tr(X_{p
\times p})+\tr(Y_{q \times q})=0$ in (\ref{bloF}).

\subsection{Cartan decompositions for
multipartite quantum systems; CCD and OED}
\label{CCDOED}

For a multipartite quantum system with $N$ subsystems of dimensions
$n_1,$...,$n_N$, the set of possible Hamiltonians is the Jordan
algebra $i\mathfrak{u}(n_1n_2 \cdots n_N)$ of $n_1n_2 \cdots n_N
\times n_1n_2 \cdots n_N$ Hermitian matrices. The Lie algebra
associated to the dynamics is $\mathfrak{u}(n_1n_2 \cdots
n_N)$. Cartan decompositions of $\mathfrak{u}(n_1n_2 \cdots n_N)$
result in decompositions of the corresponding unitary group of
quantum evolutions $U(n_1n_2 \cdots n_N)$.

The Concurrence Canonical Decomposition (CCD) was studied in
\cite{SG} \cite{SGD} as a means of decomposing the dynamics of $N$ two
level systems, into one factor which preserves the concurrence of the
density matrix, and one factor which does not. It is constructed as
follows:

Recall that the Pauli matrices
\begin{equation*}
\sigma_x=
\begin{pmatrix}
  0 & 1 \\
  1 & 0 \\
\end{pmatrix},\quad \sigma_y=
\begin{pmatrix}
  0 & -i \\
  i & 0 \\
\end{pmatrix},\quad \sigma_z=
\begin{pmatrix}
  1 & 0 \\
  0 & -1 \\
\end{pmatrix},
\end{equation*}
together with the $2\times2$ identity matrix $\1_2$, form a basis of the
Jordan algebra $i\mathfrak{u}(2)$. An orthogonal basis of
$\mathfrak{u}(2^{N})$ is given by the tensor products of the form
$i\sigma_1\otimes\cdots\otimes \sigma_N$, where
$\sigma_j=\sigma_{x,y,z}$ or $\sigma_j=\1_2$ for all $1\leq j\leq
N$. Let us denote by $i{\cal I}_o$ and ${i\cal I}_e$ the respective subspaces of
$\mathfrak{u}(2^N)$ spanned by elements of the form
$i\sigma_1\otimes\cdots\otimes \sigma_N$ with an odd or even number of factors
$\sigma_j$  given by Pauli matrices, and the remaining factors equal
to the identity $\1_2$. The CCD is the decomposition
\begin{equation}\label{CCD}
\mathfrak{u}(2^N)=i\mathcal{I}_o\oplus i \mathcal{I}_e
\end{equation}
of
$\mathfrak{u}(2^N)$. The Lie subgroup $e^{i\mathcal{I}_o}$ associated to the subalgebra
$i\mathcal{I}_o$ is a subgroup of $U({2^N})$ containing all the
local transformations. For each $L \in U(2^N)$ the decomposition
(\ref{KPDeomposition}) holds with $K \in e^{i{\cal I}_o}$  and
$P=e^{\tilde P}$ with $\tilde P \in i{\cal I}_e$. The factor $K$
and in particular, any local transformation, does not modify the
$N$-qubit concurrence \cite{SG}. Such a decomposition is of type
AII if $N$ is odd, and of type AI if $N$ is even.

\vspace{0.5cm}

The Odd-Even Decomposition (OED) was introduced in \cite{DF} as a
generalization of the CCD to multipartite systems of arbitrary
dimensions. The main idea is to construct a decomposition for the
whole Lie algebra $\mathfrak{u}(n_1 n_2 \cdots n_N)$ by combining
decompositions for the Lie algebras associated to the single
subsystems $\mathfrak{u}(n_j)$, $j=1,...,N$. This is based on the
following observation for the CCD. When writing
\begin{equation*}
\mathfrak{u}(2)=\spa\{ i \sigma_x, i\sigma_y, i\sigma_z \} \oplus
\spa\{ i\1_2 \},
\end{equation*}
we perform a (trivial) AII decomposition of $\mathfrak{u}(2)$, since
$\mathfrak{su}(2)=\mathfrak{sp}(1)$. In the CCD, we collect (modulo
$i$) tensor products with an odd number of elements in the Lie
algebra in ${\cal I}_o$ and tensor products with an even number of
elements in ${\cal I}_e$. The OED \cite{DF} is obtained by applying
this idea to general Lie algebras $\mathfrak{u}(n_j)$,
$j=1,2,\ldots, N$.  By writing
$$
\mathfrak{u}(n_j)={\cal K} \oplus {\cal P},
$$
with ${\cal K}$ conjugate to $\mathfrak{so}(n_j)$ or
$\mathfrak{sp}(\frac{n_j}{2})$, and ${\cal P}={\cal K}^\perp$,  we
obtain a decomposition of type AI or AII,  respectively. Denoting by
$\sigma$ a generic element of $i {\cal K}$ and by $S$ a generic
element of $i {\cal P}$, we define $\tilde {\cal I}_o$ and $\tilde
{\cal I}_e$ to be the respective vector space spanned by tensor products of
matrices of the type $\sigma$ and $S$ with an odd or even number of
$\sigma$ terms. The decomposition
\begin{equation}
\label{overallOED}
\mathfrak{u}(n_1 n_2 \cdots n_N)=i \tilde {\cal
I}_o \oplus i \tilde {\cal I}_e
\end{equation}
is a Cartan decomposition called the OED. The subspace $i \tilde
{\cal I}_o$ is the Lie subalgebra. This is a generalization of the
CCD not only because it applies to systems of arbitrary dimensions,
but also because, for every subsystem, we can perform different
decompositions of type AI or AII. The CCD is obtained as a special
case of the OED (\ref{overallOED}) when all the subsystems are of
dimension $2$ and a decomposition of type AII is performed on each
subsystem. Generalizing the result on the nature of the CCD
decomposition, the OED decomposition is of type AII if an odd number
of AII decompositions are performed. Otherwise, it is of type AI. As
the CCD is related to the concurrence on N qubits, the OED has the
same meaning for the generalized concurrences studied by Uhlmann in
\cite{ULL}.

We refer to \cite{BOOK} for a detailed discussion of the CCD and OED
decompositions, and to \cite{Helgason} for the mathematical
foundations of the Cartan decompositions.

\begin{remark} \label{flex} The procedure described for the OED allows one
great flexibility in the construction of various Cartan
decompositions. Not only is one free to choose decompositions of
type AI or AII for each subsystem, but one can also choose among
the different types of conjugate AI or AII decompositions for each subsystem. This
gives a method for the construction of an infinite number of
decompositions in terms of tensor product matrices, even for the
simplest case of $N$ qubits. This flexibility is crucial in  the
construction of gradings for the Lie algebra $\mathfrak{u}(n_1 n_2
\cdots n_N)$, and of recursive decompositions, as we shall see in the
following two sections. \end{remark}

\vspace{0.25cm}

We observe here that the procedure followed to construct the OED
decomposition, applying decompositions of type AI and AII, can be
used with few changes to obtain an overall decomposition starting
from  decompositions of type AIII. More specifically, we perform
decompositions of type AIII on each subsystem, and we collect in the
respective subspaces $i{\tilde {\cal I}}_o$ and $i{\tilde {\cal
I}}_e$ the linear combinations of tensor products with an odd or
even number of factors in the subalgebra part (modulo $i$).  We
again consider the decomposition of $\mathfrak{u}(n_1 n_2\cdots
n_N)$ in (\ref{overallOED}) but with $i{\tilde {\cal I}}_o$ and
$i{\tilde {\cal I}}_e$ defined in terms of type AIII decompositions.

\begin{theorem}
\label{AIIIOED} Consider the decomposition (\ref{overallOED})
obtained with decompositions of type AIII as described above. This
is a type AIII decomposition of the overall Lie algebra $u(n_1
n_2 \cdots n_N)$. If $N$ is odd, then $i\tilde {\cal I}_o$ is the Lie
subalgebra in the decomposition. If $N$ is even, then $i\tilde {\cal I}_e$
is the Lie subalgebra in the decomposition.
\end{theorem}
\begin{proof}
The proof is by induction on $N$. If $N=1$ the statement is obvious.
Assume the statement is true for $N-1$, and assume to be concrete
that $N$ is odd (exactly the same proof holds for $N$ even). Denote
by $\tilde {\cal I}^j_o$ and $\tilde {\cal I}^j_e$ the respective spaces of
matrices in $i\mathfrak{u}(n_1 n_2 \cdots n_j)$ that are linear combinations
of an odd or even number of matrices in $i {\cal K}$, where ${\cal K}$
is the subalgebra of the AIII decomposition (possibly different for
the different subsystems). Let us denote by ${\cal K}$ the
subalgebra of (block diagonal) matrices of the AIII decomposition on
the last subsystem, and by ${\cal P}$ its orthogonal complement. We
have
$$
\tilde {\cal I}_o^N=\left(\tilde {\cal I}_e^{N-1} \otimes i{\cal K}
\right) \oplus \left(\tilde {\cal I}_o^{N-1} \otimes i{\cal P}
\right), \quad \tilde {\cal I}_e^N=\left(\tilde {\cal I}_e^{N-1}
\otimes i{\cal P} \right) \oplus \left(\tilde {\cal I}_o^{N-1}
\otimes i{\cal K} \right).
$$
By the inductive assumption, there exists a unitary matrix $T$ in
$U(n_1n_2 \cdots n_{N-1})$ such that $T^\dagger \tilde{\mathcal
I}_e^{N-1}T$ is the same as the space of $n_1n_2 \cdots n_{N-1}
\times n_1n_2 \cdots n_{N-1}$ Hermitian block-diagonal matrices, and
$T^\dagger \tilde {\cal I}_o^{N-1}T$ is the same as the space of
$n_1n_2 \cdots n_{N-1} \times n_1n_2 \cdots n_{N-1}$ Hermitian
block-antidiagonal matrices. Let $T_1=T^\dagger \otimes \1_{n_N }$,
then the subspace $ T_1^\dagger \tilde {\cal I}_o^N T_1$ is spanned
by all the matrices of the form
$$
\begin{pmatrix}
A & 0 \\
0 & B
\end{pmatrix} \otimes
\begin{pmatrix}
C & 0 \\
0 & D
\end{pmatrix}\,, \text{ and }
\begin{pmatrix}
 0         &  F  \\
-F^\dagger &  0
\end{pmatrix}
\otimes
\begin{pmatrix}
 0         & G  \\
-G^\dagger & 0
\end{pmatrix}.
$$
The sizes of the matrices $A,B,C,D,F,$ and $G$ depend on the indices
$p,q$ of the two decompositions. Using the Corollary 4.3.10 of
\cite{Horn-Johnson} one can construct a permutation similarity
matrix $T_2$  so that the subspace $(T_1T_2)^\dagger \tilde {\cal
I}_o^N (T_1 T_2)$  is spanned by all the matrices of the form
$$
T_2^\dagger \begin{pmatrix}
A & 0 \\
0 & B
\end{pmatrix} \otimes
\begin{pmatrix}
C & 0 \\
0 & D
\end{pmatrix} T_2=
\begin{pmatrix}
C \otimes A & 0 & 0 & 0 \\
0 & D \otimes A & 0 & 0\\
0 & 0 & C \otimes B & 0 \\
0 & 0 & 0 & D \otimes B
\end{pmatrix}, $$
and by all the matrices of the form
$$T_2^\dagger
\begin{pmatrix}
 0         &  F  \\
-F^\dagger &  0
\end{pmatrix}
\otimes
\begin{pmatrix}
 0         & G  \\
-G^\dagger & 0
\end{pmatrix}  T_2 =
\begin{pmatrix}
0    &    0    &    0      &    G \otimes F \\
0    &    0    & -G^\dagger \otimes F & 0    \\
0    & -G\otimes F^\dagger &    0 &    0    \\
G^\dagger \otimes F^\dagger&0  & 0 &    0
\end{pmatrix}.$$
Finally, the
conjugation $P \rightarrow T_3^\dagger PT_3$, where $T_3$ has the
form
$$
T_3=\begin{pmatrix}
\1 & 0 & 0 & 0 \\
0 & 0 & 0 & \1 \\
0 & 0 & \1 & 0 \\
0 & \1 & 0 & 0
\end{pmatrix}
$$
with identity matrices $\1$ of appropriate dimensions, transforms
the subspace $ (T_1T_2)^\dagger \tilde {\cal I}_o^N (T_1T_2)$ into
the standard block diagonal form (\ref{bloF}) of the type AIII
decomposition. Therefore the subspace
$$R^\dagger \tilde {\cal I}_o^N R$$
is of form (\ref{bloF}) where $R:=T_1T_2T_3$. It can be verified
that $R$ also transforms $\tilde {\cal I}_e^N $ into the standard
block anti-diagonal form of the AIII decomposition.
\end{proof}

\begin{remark}

The indexes $p$ and $q$ of the resulting AIII decompositions of
Theorem~\ref{AIIIOED} are $p=n_1n_2\cdots n_{l-1} p_l n_{l+1} \cdots
n_N$ and $q=n_1n_2\cdots n_{l-1} q_l n_{l+1} \cdots n_N$, where $l$
may refer to any of the subsystems $l=1,...,N,$ and $p_l$ and $q_l$
are the indices of the AIII decomposition of the $l$-th system. The
theorem also indicates the inductive construction of the matrix
conjugation which maps the AIII decomposition into the standard
form. This is of interest for practical computation of the
decomposition, as most of the existing numerical algorithms refer to
the standard form (\ref{AIIIdec}), (\ref{bloF}). Note that
decompositions constructed by mixing AI or AII decompositions with
AIII-type decompositions do not give rise to Cartan decompositions.

\end{remark}

\subsection{Recursive decompositions}
\label{KGDR}
 A recursive procedure to decompose  unitary evolutions
into local and entangling factors  for the case of $N$ qubits was
introduced by N. Khaneja and S. Glaser in \cite{NS}.  One starts
with the Cartan decomposition
\begin{equation}
\label {KGD-0} \mathfrak {su}(2^N)= \mathcal K \oplus \mathcal
P
\end{equation}
of type AIII,
where
\begin{equation}
\label{KGD-I}
\begin{split}
\mathcal K = & \spa \{ \1_2\otimes A \,, \sigma_z \otimes B \, |
\,
A \in \mathfrak{su}(2^{N-1}), B \in \mathfrak{u}(2^{N-1})\} \,,\\
\mathcal P = & \spa \{ \sigma_{x,y} \otimes C \, | \, C\in \mathfrak{u}(2^{N-1})\}.\\
\end{split}
\end{equation}
This allows one to write each special unitary
evolution $L\in SU(2^N)$ as $L=K_1AK_2$, where $K_1,K_2\in
e^{\mathcal K}$ and $A\in e^{\mathcal A}$, with  $\mathcal A$
the Cartan subalgebra contained in $\mathcal P.$ The subalgebra
$\mathcal K$ is the direct sum of $\spa \{ i\sigma_z\otimes
\1_{2^{N-1}} \}$ and two copies of $\mathfrak {su}(2^{N-1})$ which
form a semisimple Lie algebra. Thus we can again apply Cartan's
Theorem to further factorize $K_1,K_2\in e^{\mathcal K}$.
This is obtained through the decomposition $\mathcal K=\mathcal
K'\oplus \mathcal P'$, with
\begin{equation}
\label{KGD-2}
\mathcal K' = \spa \{ \1_2\otimes A \, | \, A \in \mathfrak{su}(2^{N-1})\}\,,\quad
\mathcal P' =  \spa \{ \sigma_{z} \otimes B \, | \, B\in \mathfrak{u}(2^{N-1})\}\,,
\end{equation}
to decompose each $K_1\,,K_2\in e^{\mathcal K}$, thereby refining the
decomposition of $L$. The key observation is that
$\mathcal K'$ and $\mathfrak {su}(2^{N-1})$ are isomorphic,
hence the procedure can be repeated by replacing $N$ with $N-1$.

Another recursive procedure to decompose unitary evolutions was
introduced by D. D'Alessandro and R. Romano in \cite{DR}. Such a
decomposition applies to bipartite systems of arbitrary
dimensions.\footnote{Extensions to the general multipartite case can
be obtained at the price of some notational complexity.} In the first step, one starts
with an OED decomposition using AI types of decomposition on both
subsystems, so that $\mathfrak{u}(n_1n_2)$ is decomposed as in
(\ref{overallOED}), with
\begin{equation}
\label{TLC}
\tilde {\cal
I}_o:=\spa \{ \sigma \otimes S, S \otimes \sigma \} \, ,
\end{equation}
conjugate to $\mathfrak{so}(n_1n_2)$.
As $\mathfrak{so}(n_1n_2)$ is also semisimple, one then introduces a Cartan
decomposition of $\tilde {\cal I}_o$ by separating block diagonal
and anti-diagonal elements (for two arbitrary indices) in the
factors of the basis of $\tilde {\cal I}_o$. In particular, one
writes
$$
i \tilde {\cal I}_o={\cal K} \oplus {\cal P},
$$
where $${\cal K}:=\spa \{ i \sigma^D \otimes S^D, i S^D \otimes
\sigma^D, i \sigma^A \otimes S^A, i S^A \otimes \sigma^A \}$$ and
$${\cal P}:=\spa \{ i \sigma^D \otimes S^A, i S^D \otimes \sigma^A, i
\sigma^A \otimes S^D, i S^A \otimes \sigma^D \}\, ,
$$
the
superscripts $A$ and $D$ standing for block-antidiagonal and
block-diagonal respectively. The Lie algebra ${\cal K}$ is
isomorphic to the semisimple direct sum $\mathfrak{so}(r) \oplus
\mathfrak{so}(f)$ with $r+f=n_1n_2$. One
decomposes ${\cal K}$ as
$$
{\cal K}:={\cal K}' \oplus {\cal P}',
$$
with ${\cal K}':=\spa \{ i \sigma^D \otimes S^D, i S^D \otimes
\sigma^D \}$, and ${\cal P}':=\spa \{ i \sigma^A \otimes S^A, i S^A
\otimes \sigma^A \}$. The Lie subalgebra ${\cal K}'$ is isomorphic
to the direct sum of four subalgebras $\mathfrak{so}(r_1) \oplus
\mathfrak{so}(r_2) \oplus \mathfrak{so}(r_3) \oplus
\mathfrak{so}(r_4)$. Each of the summands is spanned by tensor
products of the type in (\ref{TLC}) with matrices $\sigma$ and $S$,
where only one sub-block is different from zero. One then iterates
the procedure. We refer to \cite{DR} for details.


\section{Lie algebra grading and recursive Cartan decompositions}
\label{LAG}

In this section, we give the definition of a grading of a
Lie algebra, and relate a recursive Lie algebra decomposition to a
grading. Our goal is to cast recursive decompositions of the unitary
group into a general framework. In fact, in the following section we
will show that known recursive decompositions, such as those of
Khaneja and Glaser \cite{NS} and D'Alessandro and Romano \cite{DR}
reviewed in the previous section, can be obtained from
 an appropriate grading. We shall also see in the next section
 how new decompositions can be generated with the procedure described here.

\begin{definition} Let $\mathcal L$ be a Lie algebra, and
let $M$ be an index set which has the structure of an additive semigroup. A
direct sum decomposition
$$\mathcal L=\bigoplus_{i\in M}\mathcal L_i$$
is called an $M$-\emph{grading} of $\mathcal L$ if the subspaces $\mathcal
L_i$ and $ \mathcal L_j$ satisfy the commutation relation $[
\mathcal L_i,\mathcal L_j]\subseteq \mathcal L_{i+j}$ for all
$i,j\in M$.
\end{definition}
In the special case where  $M$ is  a monoid, that is, a semigroup
with an identity element 0, the subspace $\mathcal L_0$ is a Lie
subalgebra, since it satisfies the commutation relation $[\mathcal
L_0,\mathcal L_0 ]\subseteq \mathcal L_0\,.$

\begin{example}
Consider the special linear Lie algebra $\mathfrak{sl}(2)$ of
$2\times2$ traceless matrices spanned by
$$ x = \begin{pmatrix} 0 & 1 \\0 & 0 \end{pmatrix},
\ h = \begin{pmatrix} 1 & 0 \\ 0 & -1 \end{pmatrix},\ y =
\begin{pmatrix} 0 & 0 \\1 & 0 \end{pmatrix},$$ with the commutation
relations $ [h,x]=2x\,,\ [x,y]=h\,,\ [h,y]=-2y\,.$ Let
$M=\{-1,0,1\}$. Then $M$ becomes a monoid with addition
given by the following table:
\begin{center}
\begin{tabular}{c|ccc}
\phantom{\bigg|}$(M,+)$\phantom{\bigg|}
&$\phantom{m}0\phantom{l}$
&$\phantom{l}-1\phantom{l}$
&$\phantom{l}1\phantom{l}$
\\
\hline
\phantom{\Big|}$0$
\phantom{\Big|}
 &$\phantom{m}0\phantom{l}$ &$-1$ &$1$
 \\
\phantom{\Big|}$-1$
\phantom{\Big|}
 &$\phantom{m}-1\phantom{l}$ &$0$ &$0$
 \\
\phantom{\Big|}$1$
\phantom{\Big|}
 &$\phantom{m}1\phantom{l}$ &$0$ &$0$
 \end{tabular}
\end{center}
The choice of $\mathcal L_{-1}=\spa\{x\}$, $\mathcal L_0=\spa\{h\}$,
and $\mathcal L_1=\spa\{y\}$ makes $\mathfrak{sl}(2)$ into an
$M$-graded Lie algebra.
\end{example}

\vspace{0.25cm}

A fundamental observation for what follows is that a Cartan
decomposition (\ref{CartanDecomposition}) defines a $\mathbb
Z_2$-grading of the Lie algebra $\cal L$ with ${\cal K}:= {\cal
L}_0$ and ${\cal P}:={\cal L}_1$.

As we have seen above, for  a Lie algebra $\cal L$, there are many
Cartan decompositions. The following proposition shows that $p$
Cartan decompositions give a $\mathbb Z_2^p$-grading for general
$p$.

\begin{proposition}
\label{THM-I} Consider $p$ Cartan decompositions $\mathcal
L=\mathcal L_0^j \oplus \mathcal L_1^j$ for $j=1,\ldots, p$. Define
\begin{equation}
\label{ARD-0} \mathcal L_{k_1k_2\dots k_p}:=\bigcap_{j=1,2,\ldots ,
p} \mathcal L_{k_j}^j
\end{equation}
for $k_j\in \mathbb Z_2$ and $j=1,\ldots,  p$. Then the vector space decomposition
\begin{equation}
\label{RGrading} \mathcal L=\bigoplus\mathcal L_{k_1k_2\dots k_p}
\end{equation}
forms a $\mathbb {Z}_2^{p}$-grading of $\mathcal L$.
\end{proposition}
\begin{proof}
The proof is by induction on $p$.  The claim is true for $p=1$.
Assume the claim is true for $p-1$. Let $A\in \mathcal
L_{k_1k_2\dots k_p}$ and $B\in \mathcal L_{l_1l_2\dots l_p}$ where
$k_i\,,l_i\in \mathbb Z_2$ with $1\leq i\leq p$. Then it follows
that $A\in \mathcal L_{k_1k_2\dots k_{p-1}}\,,$ $A\in \mathcal
L_{k_{p}}\,,$ $B\in\mathcal L_{l_1l_2\dots l_{p-1}}\,,$ and $B\in
\mathcal L_{l_p}$. Since $\mathcal L$ is both $\mathbb
Z_2^{p-1}$-graded and $\mathbb Z_2$-graded, we have
$$[A,B]\in  \mathcal L_{(k_1+l_1)(k_2+l_2)\dots(k_{p-1}+l_{p-1})},
\quad [A,B]\in  \mathcal L_{(k_{p}+l_{p})},$$ and
therefore
$$[A,B]\in \mathcal L_{(k_1+l_1)(k_2+l_2)\dots(k_{p-1}+l_{p-1})}
\cap \mathcal L_{(k_{p}+l_{p})}=\mathcal
L_{(k_1+l_1)(k_2+l_2)\dots(k_p+l_p)},$$ which implies that
$$ [\mathcal L_{k_1k_2\dots k_p}, \mathcal L_{l_1l_2\dots
l_p}] \subseteq \mathcal L_{(k_1+l_1)(k_2+l_2)\dots(k_p+l_p)}.$$ In
conclusion, (\ref{RGrading}) is a $\mathbb Z_2^{p}$-grading of
$\mathcal L.$
\end{proof}

Thus a Cartan decomposition of a Lie algebra is a  $\mathbb
Z_2$-grading. A combination of $p$ Cartan decompositions
gives a  $\mathbb Z_2^{p}$-grading. In order to cast a recursive
decomposition in the framework of Lie algebra gradings, we give the
following definition.

\begin{definition}
\label{recurdec}
 A \emph{recursive decomposition} of a Lie algebra ${\mathcal L}$ consists of
 two sequences of subspaces of ${\mathcal L}$, $${\cal S}_0:=\{ \mathcal
 L_0, \mathcal
 L_{00},  \mathcal
 L_{000},..., \mathcal
 L_{0^p}\}, \text{ and } {\cal S}_1:=\{ \mathcal
 L_1, \mathcal
 L_{01},  \mathcal
 L_{001},..., \mathcal
 L_{0^{p-1}1}\},$$
 both of length $p$, such that
 $$
\mathcal L_{0^j}=\mathcal L_{0^{j+1}} \oplus \mathcal L_{0^{j}1},
 $$
is a Cartan decomposition of $\mathcal L_{0^j}$ for each $j=0,...,p-1$. That is,
$$
[\mathcal L_{0^{j+1}},\mathcal L_{0^{j+1}}]\subseteq \mathcal
L_{0^{j+1}}, \quad [\mathcal L_{0^{j+1}},\mathcal
L_{0^{j}1}]\subseteq \mathcal L_{0^{j}{1}}, \quad [\mathcal
L_{0^{j}1},\mathcal L_{0^{j}1}]\subseteq \mathcal L_{0^{j+1}}.
$$
Here we have set $\mathcal L_{0^0}:=\mathcal L$ and $\mathcal
L_{0^01}:=\mathcal L_1$.
\end{definition}

\vspace{0.5cm}

Once one has a recursive decomposition of a Lie algebra $\mathcal
L$,  in the sense of the above definition,  one can obtain a
decomposition of the connected Lie group $e^{\mathcal L}$ associated
to $\mathcal L$. This is obtained by repeated use of the Cartan
decomposition theorem. Assume that $\mathcal L$ is semisimple, and
that all of the $\mathcal L_{0^j}$, $j=1,....,p-1$ are also
semisimple. One first writes each element $X$ of $e^{\mathcal L}$ as
$$
X=K_1 A K_2,
$$
where $K_1$ and $K_2$ are in $e^{\mathcal L_0}$, while $A$ belongs to
the connected Lie group corresponding to a maximal Abelian
subalgebra contained in ${\mathcal L_1}$. Then one applies the
Cartan decomposition of $\mathcal L_0$ in order to decompose $K_1$
and $K_2$, and so on. The resulting decomposition contains several
factors.

A $\mathbb Z_2^p$-grading of $\mathcal L$ induces a recursive
decomposition of $\mathcal L$ of length $p$:

\begin{proposition}
\label{grad}
 Consider a $\mathbb Z_2^p$-grading $\mathcal L=\bigoplus \mathcal R_{j_1,...,j_p}$ of $\mathcal L$. Then the sequences
${\cal S}_0:=\{{\mathcal L}_{0^k}\}$, and ${\cal S}_1:=\{{\mathcal
L}_{0^{k-1}1}\}$, defined by
$$
\mathcal L_{0^k}:=\bigoplus \mathcal R_{0^{k},j_{k+1},...,j_p}
\quad\mbox{and}\quad \mathcal L_{0^{k-1}1}:=\bigoplus \mathcal
R_{0^{k-1}1,j_{k+1},...,j_p}
$$
for $k=1,...,p$, yield a recursive decomposition of $\mathcal L$ of length $p$.
\end{proposition}

\vspace{0.5cm}

\begin{remark}
\label{aboutsemisim} Given a recursive decomposition sequence as in
Definition \ref{recurdec}, the semisimplicity of the subalgebras
${\cal L}_{0^k}$ (for $k=0,...,p$) has to be verified independently.
Even in the main case considered here, where the recursive
decomposition sequence is obtained from a Lie algebra grading by
means of combined Cartan decompositions as in Proposition
\ref{THM-I},  semisimplicity is not guaranteed. For example,  by
combining type AI and AII decompositions of $\mathfrak {su}(4)$ in
the standard basis, one obtains ${\cal L}_{00}=\mathfrak{sp}(2) \cap
\mathfrak{so}(4)$. This is not semisimple, having an element which
commutes with the whole Lie algebra. If a Lie algebra is the direct
sum of a semisimple Lie algebra and an Abelian ideal, the Cartan
decomposition theorem can be extended in the same fashion as we
extended decompositions of $\mathfrak{su}(n)$ to decompositions of
$\mathfrak{u}(n)$ in Section \ref{BMLAG}.
\end{remark}


\section{A scheme for recursive decompositions of $U(n)$}
\label{Special}

\subsection{Special cases}

We now show that the recursive decompositions of Khaneja and Glaser
\cite{NS} and D'Alessandro and Romano \cite{DR}, summarized in
Section~\ref{KGDR}, form a special case of the above
procedure. In particular, they are induced by an appropriate
grading.

Let us start with the decomposition of Khaneja and Glaser \cite{NS}.
We construct a Lie algebra grading of $\mathfrak{su}(2^N)$ using the
prescription of Proposition \ref{THM-I}. Consider $p$ Cartan
decompositions
\begin{equation}
\label{uio} \mathfrak {su}(2^N)={\cal L}_0^j \oplus {\cal L}_1^j
\end{equation}
 of $\mathfrak{su}(2^N)$ for $j=1,...,p$, all of type AIII, where ${\cal L}_0^1$
 and ${\cal L}_1^1$ are respectively equal to ${\cal L}_0$
and ${\cal L}_1$ in (\ref{KGD-I}). Now ${\cal L}_0^2$ and
${\cal L}_1^2$ are defined in the same way as ${\cal L}_0^1$ and ${\cal
L}_1^1$, except for the fact that $\sigma_x$ and $\sigma_z$ are
interchanged. Such a decomposition is conjugate to the standard type AIII decomposition,
the conjugation having the form $A \rightarrow T \otimes
{\1}_{2^{N-1}} A T^\dagger \otimes {\1}_{2^{N-1}}$ with $T$ the $2
\times 2$ matrix which diagonalizes $\sigma_x$.\footnote{There is
nothing special about $\sigma_x$ here. One could have chosen
$\sigma_y$ instead.} The summands ${\cal L}_0^3$ and ${\cal L}^3_1$ are given by
\begin{equation}
\label{KGD-Ibis}
\begin{split}
\mathcal L_0^3 = & \spa \{ A \otimes \1_2 \otimes C \,, B \otimes
\sigma_z \otimes D \, | \,
A,B \in \mathfrak{u}(2), C,D \in \mathfrak{u}(2^{N-2}), \tr(A \otimes C)=0 \} \,,\\
\mathcal L_1^3 = & \spa \{ E \otimes \sigma_{x,y} \otimes F \, | \,
E \in \mathfrak{u}(2) \,, F \in \mathfrak{u}(2^{N-2}) \}\,.
\end{split}
\end{equation}
This decomposition is again conjugate to the standard type AIII decomposition
under the permutation which  exchanges the first and second
positions. The decomposition $\mathcal L_0^4 \oplus \mathcal L_1^4$
is the same as ${\mathcal L}_0^3 \oplus \mathcal L_1^3$, except for
the fact that the roles of $\sigma_x$ and $\sigma_z$ are exchanged. The summands
${\mathcal L}_0^5$ and ${\mathcal L}_1^5$ are defined analogously to
${\mathcal L}_0^3$ and ${\mathcal L}_1^3$, using the third position
in place of the second. The same holds for ${\mathcal
L}_0^6$ and ${\mathcal L}_1^6$, defined as ${\mathcal L}_0^4$ and
${\mathcal L}_1^4$. In this fashion, one can define $p=2N-1$
decompositions\footnote{We stop at $p=2N-1$ because ${\cal L}_{0^p}$
is $\{ 0 \}.$} and therefore a $\mathbb Z_2^p$-grading of
$\mathfrak{su}(2^N)$. The corresponding pair of sequences giving the
recursive decomposition according to Proposition~\ref{grad} is
\begin{align}
\label{KGD-SQ}
\begin{split}
& {\mathcal L}_0 \,,\  {\mathcal L}_1,
 \text{ same as in (\ref{KGD-I})}, \\
& \mathcal L_{00} =   \spa \{ \1_2 \otimes A \, | A \in
\mathfrak{su}(2^{N-1}) \}\,, \\
& \mathcal L_{01} =   \spa \{ \sigma_z \otimes B \, | \, B \in
\mathfrak{u}(2^{N-1})  \}\,, \\
& \mathcal L_{000} =   \spa \{ \1_2 \otimes  \1_2 \otimes A \,, \1_2
\otimes \sigma_z \otimes C \, | \, A \in \mathfrak{su}(2^{N-2}), C
\in \mathfrak{u}(2^{N-2}) \} \,,\\
& \mathcal L_{001} =    \spa \{ \1_2 \otimes \sigma_{x,y} \otimes F
\, | \,, F \in \mathfrak{u}(2^{N-2}) \}\,,\\
&~~~~~~~~~~~~~~~~~~~~~~~~~~\vdots \\
&\mathcal L_{0^{2N-3}} =  \spa \{ {\1}_{2^{N-1}} \otimes A \,,
\1_{2^{N-2}} \otimes \sigma_z \otimes c \, | \, A \in
\mathfrak{su}(2), C \in \mathfrak{u}(2) \} \,, \\
& \mathcal
L_{0^{2N-4}1} =  \spa \{ {\1}_{2^{N-2}} \otimes \sigma_{x,y}
\otimes F \, | \,, F \in \mathfrak{u}(2) \}\,,\\
&\mathcal L_{0^{2N-2}} =  \spa \{ {\1}_{2^{N-1}}
 \otimes A  \,, A \in \mathfrak{su}(2) \} \,,\\
&
 \mathcal L_{0^{2N-3}1} = \spa \{ {\1}_{2^{N-2}}
\otimes \sigma_z \otimes F \, | \,, F \in \mathfrak{u}(2) \}\,, \\
&\mathcal L_{0^{2N-1}} = \spa \{ {\1}_{2^{N-1}} \otimes \sigma_z  \} \,,\\
& \mathcal L_{0^{2N-2}1} = \spa \{ {\1}_{2^{N-1} } \otimes
\sigma_{x,y} \}\,.
\end{split}
\end{align}
This sequence of subspaces is the one corresponding to the
Khaneja-Glaser decomposition. Notice in particular that each Lie
subalgebra $\mathcal L_{0^k}$ (for $k=1,...,{2N-1}$) is either
semisimple, or the sum of a semisimple Lie algebra and an Abelian (in
fact one-dimensional) subalgebra of elements which commute with the
whole Lie algebra. Thus the Cartan Decomposition Theorem applies in
each case (cf. Remark \ref{aboutsemisim}).

\vspace{0.5cm}

In order to obtain the recursive decomposition corresponding to the
decomposition of D'Alessandro and Romano \cite{DR}, one constructs a
grading by combining three types of decomposition:
\begin{enumerate}
\item[1)] An OED
decomposition with a type AI decomposition on each system;
\item[2)] OED
decompositions constructed using type AIII decompositions on each factor
as in Theorem~\ref{AIIIOED};
\item[3)] Type AIII decompositions in the standard
form (separating block diagonal and block antidiagonal matrices).
\end{enumerate}
In
particular, let ${\cal L}_0^1=i \tilde{\mathcal I_o}$ and ${\cal
L}_1^1=i \tilde {\cal I}_o^\perp$, where $\tilde {\cal I}_o$ is
defined in (\ref{TLC}). The summands ${\cal L}_0^2$ and ${\cal L}_1^2$ are the respective
subspaces $i \tilde {\cal I}_e$ and $i \tilde {\cal I}_o$,
referred to in Theorem~\ref{AIIIOED}. The summands ${\cal L}_0^3$
and ${\cal L}_1^3$ are the ${\cal K}$ and ${\cal P}$ subspaces of a type
AIII decomposition in standard  coordinates, with $p$ and $q$ given
by $p=p_1p_2+p_1q_2$ and $q=q_1p_2+q_1q_2$. Here $\{p_1,q_1 \}$ and
$\{p_2,q_2 \}$ are the indices for the type AIII decompositions used for
${\cal L}_0^2$ and ${\cal L}_1^2$. The summands ${\cal L}_0^4$, ${\cal L}_1^4$,
${\cal L}_0^5$ and ${\cal L}_1^5$ are constructed analogously to
${\cal L}_0^2$, ${\cal L}_1^2$, ${\cal L}_0^3$, and ${\cal
L}_1^3$ respectively, with different indices $\{p_1,q_1\}$ and
$\{p_2, q_2\}$. The same holds for ${\cal L}_0^6$, ${\cal L}_1^6$,
${\cal L}_0^7$, and ${\cal L}_1^7$, and so on. Each time, the indices $\{p_1,
q_1\}$ and $\{p_2,q_2\}$ are changed, differing
from the previous ones in order to avoid repetition of
decompositions. With these decompositions, one can define a grading,
and therefore a recursive decomposition. This decomposition
corresponds to the one in \cite{DR}.

\subsection{Construction of new recursive decompositions}
\label{CNRD}

It follows from the previous discussion that many recursive
decompositions of $\mathfrak{u}(n)$ (or $\mathfrak{su}(n)$)  and
therefore of $U(n)$ (or $SU(n)$), can be obtained. Once one has a
certain number $p$ of Cartan decompositions, then a $\mathbb
Z_2^{p}-$grading and therefore a recursive decomposition can be
obtained. We have seen that known recursive decompositions are a
special case of this general procedure. Cartan decompositions can be
obtained for example by taking one type of decomposition, e.g., AI,
and then use various conjugations. When dealing with multipartite
systems, it is convenient to have Cartan decompositions given in
terms of tensor products of matrices as the CCD and OED described in
subsection \ref{CCDOED}.

As an example we construct a new recursive decomposition of
evolutions on $N$ q-bits here. We consider the following $2N$
decompositions on $\mathfrak{u}(2^N)$.

\vspace{0.5cm}

1) A CCD decomposition so that
\begin{equation}
\mathcal L_0^1=i {\cal I}_o^N, \quad \mathcal L_1^1=i {\cal I}_e^N,
\end{equation}
where ${\cal I}_o^N$ (${\cal I}_e^N$)is the same as ${\cal I}_o$
(${\cal I}_e$) in (\ref{CCD}) with the superscript $N$ denoting the
number of positions considered;

\vspace{0.5cm}

2) An OED decomposition with all `local' decompositions of type AII
except the one on the $N-$th term which is of type AI and of the
form
$$
\mathfrak{u}(2)= \spa \{ i \sigma_z \} \oplus \spa \{ i {\bf 1_2}, i
\sigma_x,  i \sigma_y \};
$$
For the resulting decomposition, we have
\begin{equation}
\begin{split}
\mathcal L_0^2= & \spa  \{ i {\cal I}^{N-1}_e \otimes \sigma_z, \  i
{\cal I}^{N-1}_o \otimes \{\sigma_x, \sigma_y, \1_2 \}  \}\,,\\
\mathcal L_1^2= & \spa  \{ i {\cal I}^{N-1}_o \otimes \sigma_z, \ i
{\cal I}^{N-1}_e \otimes \{\sigma_x, \sigma_y, \1_2 \}  \}.
\end{split}
\end{equation}
\vspace{0.5cm}

3) Same as in 2) but with $\sigma_z$ and $\sigma_x$ interchanged to
define $\mathcal L_0^3$ and $\mathcal L_1^3$;

\vspace{0.5cm}

4) Same as in 2) but with the $N$-th position replaced by the
$N-1$-th position to define $\mathcal L_0^4$ and $\mathcal L_1^4$;

\vspace{0.5cm}

5) Same as in 4) with $\sigma_x$ and $\sigma_z$ interchanged;

\vspace{0.5cm}

6) $\rightarrow$ 2N-1) ... and so on moving toward the first
position, alternating decompositions as in 2) and decompositions as
in 3);

\vspace{0.5cm}

2N) Same as in 2) with the first position replacing the $N$-th one.

\vspace{0.5cm}

\begin{example}\label{Nuguale3} In the case $N=3$ we have, with
$\sigma$ denoting any possible Pauli matrix,
\vspace{0.5cm}
$$
\mathcal L_0^1=\spa \{i\sigma \otimes \1_2 \otimes \1_2, \,  i\1_2
\otimes \sigma \otimes \1_2, \, i\1_2 \otimes \1_2 \otimes \sigma,
\, i \sigma \otimes \sigma \otimes \sigma \}, $$
$$ \mathcal L_1^1=\spa \{i\sigma
\otimes \sigma \otimes \1_2, \, i\1_2 \otimes \sigma \otimes \sigma,
\, i\sigma \otimes \1_2 \otimes \sigma , \, i \1_2 \otimes \1_2
\otimes \1_2\},
$$
\vspace{0.25cm}
$$
\mathcal L_0^2=\spa \{i\sigma \otimes \1_2 \otimes \{\1_2,\sigma_x,
\sigma_y\}, \,  i\1_2 \otimes \sigma \otimes \{\1_2,\sigma_x,
\sigma_y\},  \, i\sigma \otimes \sigma \otimes \sigma_z , i\1_2
\otimes \1_2 \otimes \sigma_z\},
$$
$$ \mathcal
L_1^2=\spa \{i\sigma \otimes \sigma \otimes \{\1_2,\sigma_x,
\sigma_y \}, \, i\1_2 \otimes \1_2 \otimes \{\1_2, \sigma_x,
\sigma_y\}, \,  i \sigma \otimes \1_2 \otimes \sigma_z, \, i \1_2
\otimes \sigma \otimes \sigma_z \},
$$
\vspace{0.25cm}
$$
\mathcal L_0^3=\spa \{i\sigma \otimes \1_2 \otimes \{\1_2,\sigma_z,
\sigma_y\}, \,  i\1_2 \otimes \sigma \otimes \{\1_2,\sigma_z,
\sigma_y\}, \, i\sigma \otimes \sigma \otimes \sigma_x , i\1_2
\otimes \1_2 \otimes \sigma_x\},
$$
$$ \mathcal
L_1^3=\spa \{i\sigma \otimes \sigma \otimes \{\1_2,\sigma_z,
\sigma_y \}, \, i\1_2 \otimes \1_2 \otimes \{\1_2, \sigma_z,
\sigma_y\}, \,  i \sigma \otimes \1_2 \otimes \sigma_x, \, i \1_2
\otimes \sigma \otimes \sigma_x \},
$$
$$
\vdots
$$
$$
\mathcal L_0^6=\spa \{i\{\1_2, \sigma_x, \sigma_y\}\otimes \sigma
\otimes \1_2 , \, i\{\1_2, \sigma_x, \sigma_y\}  \otimes \1_2
\otimes \sigma, \, i \sigma_z \otimes \sigma \otimes \sigma, \, i
\sigma_z \otimes \1_2 \otimes \1_2 \},
$$
$$ \mathcal L_1^6=\spa \{i \{ \1_2,
\sigma_x, \sigma_y \} \otimes \sigma \otimes \sigma,  i \{
\1_2,\sigma_x, \sigma_y\} \otimes \1_2 \otimes \1_2, \,  i \sigma_z
\otimes \sigma \otimes \1_2, \, i \sigma_z \otimes \1_2 \otimes
\sigma\}.
$$
\end{example}

\vspace{0.25cm}

In the general case, with the decompositions
$\mathfrak{u}(2^N):=\mathcal L_0^j \oplus \mathcal L_1^j$,
$j=1,...,2N$ one  constructs a grading as in Proposition \ref{THM-I}
and a recursive decomposition according to Proposition \ref{grad}.
The sequences of subspaces associated to the latter are given,   for
$k=0,\ldots,N-1,$\footnote{If the factors on the left occupy all the
$N$ positions in the tensor products, the factors on the right do
not appear.}
\begin{equation*}
\begin{split}
&\mathcal L_{0^{2k+1}}= \spa \{ i {\cal I}^{N-k}_o \otimes \1_{2^k} \}, \\
&\mathcal L_{0^{2k}1} = \spa \{ i {\cal I}_e^{N-k} \otimes \sigma_z \otimes \1_{2^{k-1}} \},\\
&\mathcal L_{0^{2k+2}}= \spa \{ i {\cal I}_o^{N-k-1} \otimes
\1_{2^{k+1}}, \, i {\cal I}_e^{N-k-1} \otimes
\sigma_z \otimes \1_{2^k} \},\\
& \mathcal L_{0^{2k+1} 1} = \spa \{ i {\cal I}_e^{N-k-1} \otimes
\{\sigma_x, \sigma_y\} \otimes \1_{2^k} \}.
\end{split}
\end{equation*}

In order to apply this recursive decomposition for the recursive
decomposition of the Lie group $U(2^N)$ we make the following two
remarks.

\begin{remark}\label{REM1}

The Lie subalgebra  $\mathcal L_{0^{2k+1}}= \spa \{ i{\cal
I}_o^{N-k} \otimes \1_{2^{N-k}}\}$, $ 0 \leq k \leq N-1$, is
isomorphic to $i{\cal I}_o^{N-k}$ which is conjugate to $\mathfrak
{so}(2^{N-k})$ or $\mathfrak {sp}(2^{N-k-1})$ according to whether
${N-k}$ is even or odd, respectively. Thus, in every case the Lie
algebra is semisimple. On the other hand, the Lie subalgebra
$\mathcal L_{0^{2k}}=\spa \{ i {\cal I}_o^{N-k}\otimes \1_{ 2^k } ,
\, i {\cal I}_e^{N- k} \otimes \sigma_z \otimes \1_{ 2^{k-1} } \}$
is isomorphic to $\mathfrak{u}(2^{N-k})$, and the isomorphism is
given by the map
\begin{equation}
\label{I9} A \otimes \1_{2^k} \longmapsto A\, ,\quad B\otimes
\sigma_z \otimes \1_{2^{k-1}}\longmapsto B\, ,
\end{equation}
where $A\in i\mathcal I_o^{N-k}$ and $B\in i\mathcal I_e^{N-k}$.
This is a the direct sum of a semisimple Lie algebra and a one
dimensional subspace whose elements all commute with the elements of
the  Lie algebra. In all cases, the Cartan decomposition theorem
applies.
\end{remark}

\vspace{0.25cm}

\begin{remark}\label{rank}
In applying Cartan theorem to obtain a $KAK$ decomposition as in
(\ref{KAK2}) we need to identify the rank and a Cartan subalgebra at
each step. The decomposition
$$
\mathcal L_{0^{2k}}=\mathcal L_{0^{2k+1}} \oplus \mathcal
L_{0^{2k}1},
$$
with $k=0,...,N-1$ is a decomposition of type AI or AII (modulo the
isomorphism in (\ref{I9})) of $\mathfrak{u}(2^{N-k})$, according
whether $N-k$ is even or odd, respectively. In the AI case the rank
is $2^{N-k}$. A maximal Abelian subalgebra is spanned by the
subspace
$$
 {\cal H}_{AII}=\spa \{ i {\cal H}_{\frac{{N-k}}{2}}
 \otimes \sigma_z \otimes \1_{2^{k-1} } \}.
$$
Here we have used the following notation
\begin{equation}\label{acca} \mathcal{H}:=\spa \{ \sigma_x\otimes\sigma_x \,,
\sigma_y\otimes\sigma_y \,, \sigma_z\otimes\sigma_z \,,
\1_2\otimes\1_2 \},
\end{equation}
and $\mathcal{H}_l$ denotes the set obtained by  tensor products of
$l$ elements  of $\mathcal{H}$, that is,
$\mathcal{H}_l=\mathcal{H}\otimes\cdots\otimes \mathcal{H},$ $l$
times.\footnote{Using the fact that ${\cal H}$ is a commuting set
and induction on $l$ along with the formula $$ [K\otimes L, M\otimes
N]=[K,M]\otimes \{L,N\}+\{K,M\}\otimes[L,N]\, ,
$$
 it is easy to see that $\mathcal{H}_l$ is also a  commuting set.}
In the odd, AII, case, the rank is $2^{N-k-1}$. A Cartan subalgebra
in this case is given by
 \begin{equation*}
{\cal H}_{AII}:=\spa  \{i\mathcal H_{\frac{N-k-1}{2}} \otimes \1_2
\otimes\, \sigma_z \otimes \1_{2^{k-1} } \}\,.
\end{equation*}
\vspace{0.25cm}
 The decomposition
$$
\mathcal L_{0^{2k+1}}=\mathcal L_{0^{2k+2}} \oplus \mathcal L_{0^{2k+1}1},
$$
with $k=0,...,N-1$, is a decomposition of $\mathfrak {so}(2^{N-k})$
or $\mathfrak {sp}(2^{N-k-1})$ according to whether ${N-k}$ is even
or odd. In the first case, it is a decomposition of type DIII (we
refer to \cite{Helgason} for decompositions of Lie algebras
different from $\mathfrak{u}(n)$) which has rank $2^{N-k-2}$. The
Cartan subalgebra can be taken equal to
\begin{equation*}
{\mathcal H}_{DIII}:=\spa \{i {\cal H}_{\frac{N-k-2}{2}} \otimes
\1_2 \otimes\,  \sigma_x \otimes \1_{2^k} \}.
\end{equation*}
In the second case, it is a decomposition of type CI and the
associated rank is $2^{N-k-1}$. The Cartan subalgebra can be taken
equal to
\begin{equation*}
{\mathcal H}_{CI}:=\spa \{i {\cal H}_{\frac{N-k-1}{2}}  \otimes\,
\sigma_x \otimes \1_{2^k} \}.
\end{equation*}
\end{remark}


\section{An Example of Computation}
\label{CIE}

In this section, we use an example to discuss some of the
computational issues arising in recursive decompositions. In
particular, we focus on the  application of the recursive procedure
described in the previous section to a generalized SWAP operator
$X_{sw}\in U(8)$. In the tensor product basis, the action of
$X_{sw}$ is defined by
\begin{equation*}
X_{sw} :|i\rangle\otimes|j\rangle\otimes|k\rangle \mapsto
|j\rangle\otimes|k\rangle\otimes|i\rangle\, ,
\end{equation*}
where $i,j,k=0,1\, ,$ refers to an orthonormal basis $\{\,
|0\rangle,|1\rangle \}$ of the Hilbert space of each of three two
level  systems.  $X_{sw}$ is the cyclic left shift operator acting
on three qubits.  The matrix representation of this operator is
given by
\begin{equation*}
X_{sw}={\footnotesize\begin{pmatrix}
     1  &   0  &   0  &   0  &   0  &   0  &   0  &   0\\
     0  &   0  &   1  &   0  &   0  &   0  &   0  &   0\\
     0  &   0  &   0  &   0  &   1  &   0  &   0  &   0\\
     0  &   0  &   0  &   0  &   0  &   0  &   1  &   0\\
     0  &   1  &   0  &   0  &   0  &   0  &   0  &   0\\
     0  &   0  &   0  &   1  &   0  &   0  &   0  &   0\\
     0  &   0  &   0  &   0  &   0  &   1  &   0  &   0\\
     0  &   0  &   0  &   0  &   0  &   0  &   0  &   1\\
\end{pmatrix}}.
\end{equation*}
Our goal is to factorize $X_{sw}$ in terms of elementary matrices.
Using the Cartan decomposition of the previous section  (cf. example
\ref{Nuguale3}), one can construct a grading and therefore obtain a
recursive decomposition of $\mathfrak{u}(8)$. Modulo isomorphisms,
the  sequences characterizing the recursive decomposition are given
by
\begin{equation}\label{ADD1}
\mathcal S_0=\{
\mathfrak{sp}(4),\mathfrak{u}(4),\mathfrak{so}(4),\mathfrak{u}(2),\mathfrak{sp}(1),\mathfrak{u}(1)
\},\ \  \mathcal S_1=\{\mathfrak{sp}(4)^\perp
,\mathfrak{u}(4)^\perp,\mathfrak{so}(4)^\perp,\mathfrak{u}(2)^\perp,\mathfrak{sp}(1)^\perp,\mathfrak{u}(1)^\perp
\}. \end{equation}
 Most of the algorithms for the computation of decompositions of the unitary group are given in standard
coordinates. To transform the problem into standard coordinates, one
uses an orthogonal change of basis. According to \cite{SG} the
associated matrix is given by
\begin{equation}\label{ADD2}
F={\footnotesize\frac{1}{\sqrt2}\begin{pmatrix}
   1   &      0 &      0 &      0 &      1 &      0 &      0 &      0\\
   0   &      1 &      0 &      0 &      0 &     -1 &      0 &      0\\
   0   &      0 &      1 &      0 &      0 &      0 &     -1 &      0\\
   0   &      0 &      0 &      1 &      0 &      0 &      0 &      1\\
   0   &      0 &      0 &      1 &      0 &      0 &      0 &     -1\\
   0   &      0 &      1 &      0 &      0 &      0 &      1 &      0\\
   0   &      1 &      0 &      0 &      0 &      1 &      0 &      0\\
   1   &      0 &      0 &      0 &     -1 &      0 &      0 &      0\\
\end{pmatrix}}.
\end{equation}
This matrix is referred to as the \emph{finagler}. After this change
of coordinates,  $X_{sw}$ takes the form  $\tilde{X}_{sw}=F^T
X_{sw}F$, with  $\tilde{X}_{sw}= \1_2\otimes X'_{sw}$ where
\begin{equation*}
X'_{sw}={\footnotesize\begin{pmatrix}
     1  &   0  &   0  &   0  \\
     0  &   0  &   1  &   0  \\
     0  &   0  &   0  &   1  \\
     0  &   1  &   0  &   0  \\
\end{pmatrix}}\,.
\end{equation*}
To perform the decomposition, we follow the sequence of subspaces in
(\ref{ADD1}).  The first step is to compute the decomposition of
$\tilde{X}_{sw}$ induced by the Cartan pair
$(\mathfrak{sp}(4),\mathfrak{sp}(4)^\perp)$ of $\mathfrak{u}(8)$. It
can be verified that $\tilde{X}_{sw}$ is symplectic, i.e.,
$\tilde{X}_{sw}\in Sp(4)$, therefore its decomposition is trivial.
Moreover, $\tilde{X}_{sw}$ is contained in the image of $U(4)$
embedded into $Sp(4)$,\footnote{An embedding of $ U(n)$ into $Sp(n)$
or $SO(2n)$ is given by the map $U+iV \mapsto \left(
  \begin{smallmatrix}
    U & V \\
    -V & U \\
  \end{smallmatrix}
\right)$ where $U$ and $V$ are real matrices.} and represented by
$X'_{sw}$ in $U(4)$. Indeed, $X'_{sw}$ is not only unitary but
orthogonal, i.e., $X'_{sw}\in SO(4)$. Hence the decompositions
induced by the Cartan pairs
$(\mathfrak{u}(4),\mathfrak{u}(4)^\perp)$, and
$(\mathfrak{so}(4),\mathfrak{so}(4)^\perp)$ are also trivial. The
computational problem is now to find the decomposition
$X'_{sw}=K'_1A'K'_2$ induced by the Cartan pair
$(\mathfrak{u}(2),\mathfrak{u}(2)^\perp)$ so that $K'_1$ and $K'_2$
are contained the image of $U(2)$ embedding into $SO(4)$, and $A'$
is the exponential of an element of the suitable Cartan subalgebra,
i.e., $A'= \left(
  \begin{smallmatrix}
    E & 0 \\
    0 & E^{-1} \\
  \end{smallmatrix}
\right).$ Let us partition $X'_{sw}$ into $2\times 2$ blocks, i.e.,  $$X'_{sw}= \left(
  \begin{matrix}
    X_{11} & X_{12} \\
    X_{21} & X_{22} \\
  \end{matrix}
\right).$$ Choose $K_2'=\1_4$.  Then $X'_{sw}$ decomposes as
\begin{equation}
\label{meq}
X'_{sw}=
\begin{pmatrix}
A & B \\ -B & A
\end{pmatrix}
\begin{pmatrix}
E & 0 \\ 0 & E^{-1}
\end{pmatrix},
\end{equation}
where $A+iB \in U(2)$. This equation is equivalent to two matrix equations
$$
X_{11}-iX_{21}=(A+iB)E,\quad X_{22}+iX_{12}=(A+iB)E^{-1},\quad
$$
which implies that
$$E^2=(X_{22}+iX_{12})^{-1}(X_{11}-iX_{21})=\begin{pmatrix} 0 &-1 \\1 & 0 \end{pmatrix}.$$
Once $E$ is determined from the last equation, we obtain $A$ and $B$ using (\ref{meq}) so that
\begin{equation}\label{ADD3} K_1'={\footnotesize\frac{1}{\sqrt
2}\begin{pmatrix}
     1  &   1  &   0  &   0  \\
     0  &   0  &   1  &  -1  \\
     0  &   0  &   1  &   1  \\
     -1 &   1  &   0  &   0  \\
\end{pmatrix}},\quad
A'={\footnotesize \frac{1}{\sqrt 2}
\begin{pmatrix}
     1  &  -1  &   0  &   0  \\
     1  &   1  &   0  &   0  \\
     0  &   0  &   1  &   1  \\
     0  &   0  &  -1  &   1  \\
\end{pmatrix}}.
\end{equation}
In the next step, we decompose $K_1'$ using the Cartan pairs
$(\mathfrak{sp}(1),\mathfrak{sp}(1)^\perp)$ and
$(\mathfrak{u}(1),\mathfrak{u}(1)^\perp)$.

As final result, we obtain
\begin{equation}
\label{tswap}
\tilde{X}_{sw}=\tilde{L}_1 \tilde{L}_2\tilde{L}_3 \tilde{L}_4,
\end{equation}
where
\begin{align*}
& \tilde{L}_1  = \frac{1}{\sqrt 2} (\1_8-i\1_2\otimes \sigma_y\otimes \1_2)\,, \quad
\tilde{L}_2  = \frac{1}{\sqrt 2} (\1_8+i\1_2\otimes \sigma_y\otimes \sigma_z) \,,\\
& \tilde{L}_3  = \frac{1}{\sqrt 2} (\1_8 + i\1_4 \otimes \sigma_y )\,,\qquad\quad~
 \tilde{L}_4  = \1_2\otimes A',
\end{align*}
where $A'$ is defined in (\ref{ADD3}). We map $\tilde{X}_{sw}$ in
(\ref{tswap}) back to the tensor product basis to write
\begin{equation}\label{CIE-2}
X_{sw}=L_1L_2L_3L_4,
\end{equation}
where $L_k= F\tilde{L}_k F^T$, $1\leq k\leq 4$, where $F$ is the
finagler defined in (\ref{ADD2}).  Finally, we write all the factors
in (\ref{CIE-2}) as exponentials of matrices in the tensor product
basis to obtain
\begin{equation}
\label{swap1} X_{sw}=\
e^{\frac{-i\pi}{4}\sigma_y\otimes\sigma_z\otimes\sigma_x}\,
e^{\frac{i\pi}{4}\sigma_x\otimes\sigma_z\otimes\sigma_y}\,
e^{\frac{i\pi}{4}\sigma_y\otimes\sigma_x\otimes\sigma_z}\,
e^{\frac{-i\pi}{4}\sigma_x\otimes\sigma_y\otimes\sigma_z}\, .
\end{equation}

For sake of comparison, we factorize $X_{sw}$ using the
decomposition of Khaneja and Glaser \cite{NS}. We have shown that
this factorization corresponds to the sequences
$$\mathcal S_0=\{ \mathcal L_{0},\ \mathcal L_{0^2},\ \mathcal L_{0^3},\ \mathcal L_{0^4}, \mathcal L_{0^5} \}\,,\quad
\mathcal S_1=\{ \mathcal L_{1},\ \mathcal L_{01},\ \mathcal
L_{0^21},\ \mathcal L_{0^31},  \mathcal L_{0^41}\} $$ (cf. the
elements of  $\mathcal S_0$ and $\mathcal S_1$ in (\ref{KGD-SQ})
with $N=3$). Recall that $\mathcal L_0=\spa\{ \1\otimes A, \sigma_z
\otimes B\, |\, A\in \mathfrak{su}(4), B\in \mathfrak{u}(4)\}$. We
find convenient to choose the Cartan subalgebra as the span of
matrices of type $\sigma_x\otimes D$ with $D$ diagonal for the
Cartan pair $(\mathcal L_0,\mathcal L_1)$ of $\mathfrak{u}(8)$.
Therefore the corresponding decomposition of $X_{sw}$ is given by
$$X_{sw} = K_1 A K_2,$$
where $K_j=diag (K_{j1},K_{j2})$, with
$K_{jk}$, $1\leq j,k\leq 2$, $4\times 4$ unitary and $A= \left(
\begin{smallmatrix} D_1 & D_2 \\
 D_2 & D_1
\end{smallmatrix}\right)
$ where $D_j$ diagonal with $D_1^2-D_2^2=\1_4$. Following the
procedure described in \cite{BOOK} (section 8.2.3), we obtain the
matrices
$$K_{11} = \begin{pmatrix}
     1  &   0  &   0  &   0  \\
     0  &   0  &   0  &   1  \\
     0  &  -1  &   0  &   0  \\
     0  &   0  &  -1  &   0  \\
\end{pmatrix},\ \ K_{12} = \begin{pmatrix}
     0  &   1  &   0  &   0  \\
     0  &   0  &  -1  &   0  \\
    -1  &   0  &   0  &   0  \\
     0  &   0  &   0  &   1  \\
\end{pmatrix},\ \ K_{21} = \begin{pmatrix}
     1  &   0  &   0  &   0  \\
     0  &  -i  &   0  &   0  \\
     0  &   0  &   0  &  -i  \\
     0  &   0  &   1  &   0  \\
\end{pmatrix}, $$

$$ K_{22} = \begin{pmatrix}
     0  &  -1  &   0  &   0  \\
     i  &   0  &   0  &   0  \\
     0  &   0  &  -i  &   0  \\
     0  &   0  &   0  &   1  \\
\end{pmatrix},\quad  D_{1} = \begin{pmatrix}
     1  &   0  &   0  &   0  \\
     0  &   0  &   0  &   0  \\
     0  &   0  &   0  &   0  \\
     0  &   0  &   0  &   1  \\
\end{pmatrix},\quad  D_{2} = \begin{pmatrix}
     0  &   0  &   0  &   0  \\
     0  &   i  &   0  &   0  \\
     0  &   0  &  -i  &   0  \\
     0  &   0  &   0  &   0  \\
\end{pmatrix}.
$$

The next step is to factorize both $K_1$ and $K_2$ using the Cartan
pair $(\mathcal L_{0^2},\mathcal L_{01})$. Notice that $\mathcal
L_{0}^2$ is given by $\1\otimes \mathfrak{su}(4)$. Choosing the
Cartan subalgebra as the span of matrices of type $\sigma_z\otimes
D$, with diagonal $D$, induces the decomposition
\begin{equation}
\label{KG2DEC}
\begin{pmatrix}
K_{j1} & 0\\
0 &K_{j2}
\end{pmatrix}
=
\begin{pmatrix}
L_{j1} & 0\\
0 &L_{j1}
\end{pmatrix}
\begin{pmatrix}
A_j & 0\\
0 & A_j^{-1}
\end{pmatrix}
\begin{pmatrix}
L_{j2} & 0\\
0 &L_{j2}
\end{pmatrix}
\end{equation}
where $L_{j1},L_{j_2}\in SU(4)$ and $A_j$ diagonal.
In order to achieve this decomposition, we set
$$ \begin{pmatrix}
K_{j1} & 0\\
0 &K_{j2}
\end{pmatrix}
=
\begin{pmatrix}
K & 0\\
0 & K
\end{pmatrix}
\begin{pmatrix}
P & 0\\
0 & P^\dagger
\end{pmatrix}$$ with unitary $K$ and $P$ to obtain two matrix equations $K_{j1}=KP$ and $K_{j2}=KP^\dagger$. Then it follows that $P^2=K_{j2}^\dagger K_{j1}$. We diagonalize $P^2$ with a unitary matrix $U$ to write $P^2=U \Lambda U^\dagger$, and we choose $D=\Lambda^{\frac{1}{2}}$ with $det(D)=1$ so that $P=U D U^\dagger$. Once $P$ is determined,
$K$ can be found from the matrix equation $K_{j2}=KP^\dagger$. Finally we choose $L_{j1}=KU$, $L_{j2}=U^\dagger$ and $A_j=D$ to obtain the desired decomposition (\ref{KG2DEC}).

Applying this procedure, we obtain
$$ L_{11} = \frac{1}{\sqrt2}
\begin{pmatrix}
i  &  0  &  0  &  1  \\
0  & -i  &  1  &  0  \\
i  &  0  &  0  & -1  \\
0  &  i  &  1  &  0 \end{pmatrix},\ \
L_{12} = \frac{1}{\sqrt2}
\begin{pmatrix}
-1 &  1  &  0  &  0  \\
 0 &  0  & -1  & -1  \\
 0 &  0  & -1  &  1  \\
 1 &  1  &  0  &  0 \end{pmatrix},\ \
L_{21} = \frac{1}{\sqrt2}
\begin{pmatrix}
i  &  0  &  0  & -1  \\
1  &  0  &  0  & -i  \\
0  &  1  & -i  &  0  \\
0  & -i  &  1  &  0 \end{pmatrix} ,$$
$$
 L_{22} = \frac{1}{\sqrt2}
 \begin{pmatrix}
-1 & -1  &  0  & 0  \\
 0 &  0  & -1  & 1  \\
 0 &  0  &  1  & 1  \\
-1 &  1  &  0  & 0 \end{pmatrix},\quad A_1 = A_2 = \begin{pmatrix}
i  &  0  &  0  & 0  \\
0  & -i  &  0  &0  \\
0  &  0  &  1  &0  \\
0  &  0  &  0  &1 \end{pmatrix} .
$$
Similarly, we repeat the first two steps with the respective Cartan
pairs $(\mathcal L_{0^3},\mathcal L_{0^21})$ and $(\mathcal
L_{0^4},\mathcal L_{0^31})$ to decompose $L_{jk}$, with $ 1\leq j,k
\leq 2$. Finally, writing  all the factors as exponentials,  we
obtain the factorization
\begin{equation}
\label{swap2} X_{sw}=K_1AK_2,
\end{equation}
with
\begin{equation*}
\begin{split}
K_1 = \ & e^{\frac{i\pi}{4}\II\ot \Iz}
e^{\frac{i\pi}{4}\I\ot\Iz\ot\Iz} e^{\frac{i\pi}{4}\I\ot\Ix\ot\I}
e^{\frac{-i\pi}{4}\II\ot \Iy} e^{\frac{i\pi}{4}\II \ot \Ix}
e^{\frac{i\pi}{4}\I \ot\Iz\ot\Iz} e^{\frac{-i\pi}{4}\II\ot\Ix}\\
& e^{\frac{i\pi}{4}\Iz\ot\I\ot\Iz} e^{\frac{i\pi}{4}\Iz\ot\Iz\ot\Iz}
e^{\frac{i\pi}{4}\I\ot\Ix\ot\I} e^{\frac{-i\pi}{4}\I\ot\Ix\ot\Iz}
e^{\frac{i\pi}{4}\I\ot\Iz\ot\I} e^{\frac{-i\pi}{4}\I\ot\Iz\ot\Iz}
e^{\frac{i3\pi}{4}\II\ot\Iy},
\end{split}\end{equation*}
$$ A = \ e^{\frac{i\pi}{4}\Ix\ot\Iz\ot\I}
e^{\frac{-i\pi}{4}\Ix\ot\I\ot\Iz},$$
\begin{equation*}
\begin{split}
K_2 = \ & e^{\frac{i\pi}{2}\II\ot\Iy}
e^{\frac{-i\pi}{4}\I\ot\Iz\ot\I} e^{\frac{i\pi}{4}\I\ot\Iz\ot\Iz}
e^{\frac{i3\pi}{4}\II\ot\Iy} e^{\frac{i\pi}{4}\I\ot\Ix\ot\I}
e^{\frac{-i\pi}{4}\I\ot\Ix\ot\Iz} e^{\frac{i\pi}{4}\Iz\ot\I\ot\Iz}
e^{\frac{i\pi}{4}\Iz\ot\Iz\ot\Iz}\\
& e^{\frac{i\pi}{2}\II\ot\Iz} e^{\frac{i\pi}{2}\I\ot\Iz\ot\Iz}
e^{\frac{i\pi}{4}\I\ot\Ix\ot\I} e^{\frac{-i\pi}{4}\I\ot\Ix\ot\Iz}
e^{\frac{i\pi}{2}\II\ot\Iy} e^{\frac{-i\pi}{4}\I\ot\Iz\ot\I}
e^{\frac{-i\pi}{4}\I\ot\Iz\ot\Iz} e^{\frac{-i\pi}{4}\II\ot\Iy},
\end{split}\end{equation*}
where we used $\1$ to denote $\1_2$.

\section{Conclusions}
\label{conc}

Grading of a Lie algebra, Cartan decompositions and recursive
decompositions of a Lie group are interrelated ideas. From a set of
$p$ Cartan  decompositions,  one can naturally obtain a $\mathbb
Z_2^{p}-$grading  of a Lie algebra and a recursive decomposition of
the associated Lie group. Known procedures for the recursive
decomposition of the unitary group of quantum evolutions are special
cases of this general scheme. When dealing with multipartite quantum
systems, it is convenient if   the decompositions used in the
procedure are given in terms of tensor products of basis elements of
the Lie algebras associated to the single subsystems. This is the
case for the CCD decomposition on $n$-qubits and the OED
decomposition in its various forms. In this way, the factors of each
element of the group are exponentials of tensor products, and one
can identify local operations as well as multi-body interactions.

We have given a new recursive decomposition applying the general
procedure, along with an example of computation (section \ref{CIE}).
For this example , formulas (\ref{swap1}) and (\ref{swap2}), which
is obtained applying the results of \cite{NS},  give different
decompositions. In general, different recursive decompositions of
$\mathfrak {u}(n)$ will result in different factorizations of
$U(n)$. The framework presented here gives a virtually unbounded
number of alternatives to decompose $U(n)$ and parametrize quantum
evolutions.



\end{document}